\documentclass[9pt,twocolumn,twoside]{opticajnl}
\journal{opticajournal} 

\setboolean{shortarticle}{false}

\usepackage{lineno}

\title{Burst-mode timing recovery based on fourth-power phase detector for passive optical networks}

\author[1,~*]{Ji Zhou}
\author[2]{Haide Wang}
\author[1]{Xiaofeng Zhang}
\author[1]{Zhiyang Liu}
\author[3]{Miao Yu}
\author[4]{Changyuan Yu}
\author[1,~5]{Liangchuan Li}
\author[1,~6]{Xiangjun Xin}

\affil[1]{Aerospace and Informatics Domain, Beijing Institute of Technology, Zhuhai 519088, China}
\affil[2]{School of Cyber Security, Guangdong Polytechnic Normal University, Guangzhou 510665, China}
\affil[3]{China Mobile Communications Corporation Group Co. Ltd, Beijing, China}
\affil[4]{Department of Electrical and Electronic Engineering, The Hong Kong Polytechnic University, Hong Kong, China}
\affil[5]{liliangchuan@bitzh.edu.cn}
\affil[6]{xinxiangjun@bit.edu.cn}
\affil[*]{zhouji@bitzh.edu.cn}

\begin{abstract}
Driven by the ever-increasing capacity demands, 50G passive optical network (50G-PON) is ready for practical application. It is highly challenging to realize 50GHz burst-mode analog components; therefore, based on 25GHz burst-mode analog devices, burst-mode digital signal processing (DSP) is introduced to achieve the reception and processing of 50Gb/s on-off keying burst signals. To optimize the power consumption and area of the DSP chip, a one-sample-per-symbol (1-SPS) analog-to-digital converter has been applied in 50G-PON. One of the main challenges is implementing burst-mode timing recovery (BM-TR) for the 1-SPS burst signal in 50G-PON. In this paper, we first propose a BM-TR based on the fourth-power phase detector (4PPD) for 50G-PON. We mathematically verify that 4PPD can directly compute the timing phase offset (TPO) from the 1-SPS signal without using training sequences, allowing for immediate BM-TR initialization within 20 cycles to prevent long convergence times. After the initialization, the feedback loop structure tracks the TPO changes based on the sign of 4PPD, the loop filter, and the numerically controlled oscillator. In conclusion, training-sequence-free 4PPD-based BM-TR achieves low burst overhead via fast convergence and is particularly effective for handling burst signals in 50G-PON.
\end{abstract}

\setboolean{displaycopyright}{false} 
\doi{}
\begin{document}

\maketitle

\section{Introduction}
\begin{figure}[!t]
\centering
\includegraphics[width=\linewidth]{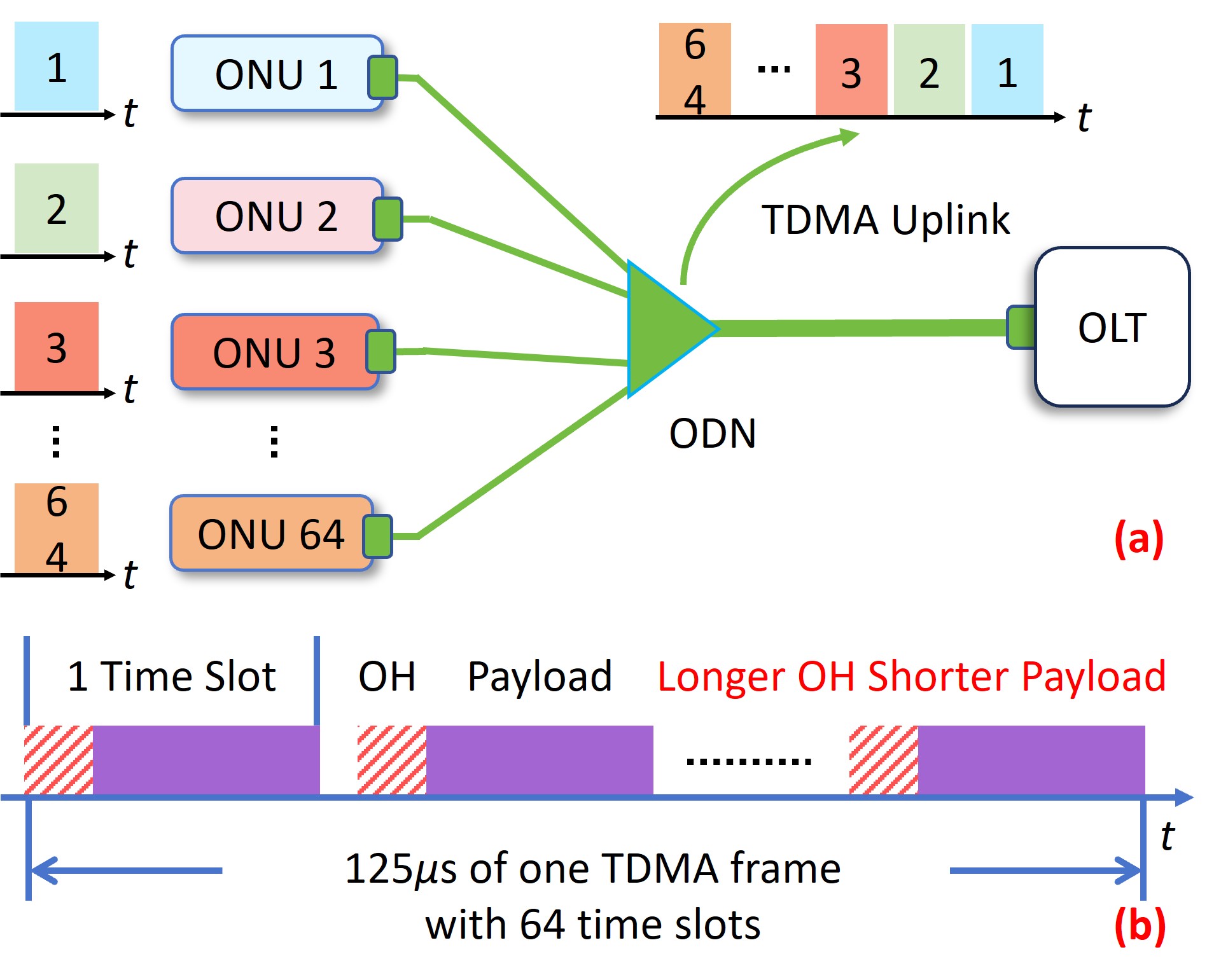}
\caption{(a) The TDMA uplink architecture of the PON. (b) The 125$\mu$s burst frame structure of the TDMA with 64 time slots. OH: overhead.}
\label{fig:1}
\end{figure}

The advancement of passive optical networks (PON) is fueled by artificial intelligence, mobile internet, and high-definition video \cite{jia2025coherent, 11020587, effenberger2024new}. The International Telecommunication Union Telecommunication Standardization Sector (ITU-T) has standardized the G.9804 series, known as Higher Speed PON, to define 50G-PON capable of delivering up to 50 Gbit/s line rates, addressing the growing demand for higher network capacity \cite{zhang2020progress, bonk202250g, zhang2021carrier, zhou2026higher}. Digital signal processing (DSP) has been introduced to compensate for inter-symbol interference caused by the limited bandwidth and chromatic dispersion in 50G-PON, distinguishing it from previous PON generations \cite{li2020dsp, liang2022dsp, torres2022overview}. Since the beginning of commercial PONs, statistical multiplexing via time-division multiple access (TDMA) ensures capacity and quantity for subscribers \cite{chung2022tdm, dhaini2013energy, harstead2018technology}. Fig. \ref{fig:1} (a) shows the TDMA uplink architecture of the PON. The optical network units (ONUs) transmit their burst signal in the allocated time slots, which are then passively aggregated through the optical distribution network (ODN) to the optical line terminal (OLT). The OLT should handle the dynamic distortions on the burst signals from the ONUs. In previous PON generations, analog burst-mode timing recovery (BM-TR) was used to process the TDMA burst signal \cite{shastri20105, su201410, lin20142}. Therefore, for the DSP-enabled 50G-PON, the digital BM-TR is crucial for processing high-speed burst signals and determining product competitiveness \cite{zhou2025real, zhang202550gb, ozkaya201856gb}.

To reduce the power consumption and area of the DSP chip, 50G-PON typically employs a one-sample-per-symbol (1-SPS)  sampling analog-to-digital converter (ADC) \cite{kaneda2020dsp}. For a 1-SPS digital signal, timing recovery based on the MM phase detector (MM-TR) and feedback loop structure gradually converges to the right timing phase offset (TPO) \cite{mueller1976timing}. The MM phase detector can only estimate the trend, not the precise value, of the TPO. The feedback loop structure, based on a loop filter and a numerically controlled oscillator (NCO), gradually approaches the correct TPO by utilizing the trend of the TPO, which necessitates a long convergence time. Fig. \ref{fig:1} (b) depicts the burst frame structure of the TDMA. One TDMA frame has 64 time slots with a 125$\mu$s duration for up to 64 ONUs. Due to the fixed 125$\mu$s duration, the long convergence time increases the overhead and decreases the payload. Meanwhile, the MM-TR struggles to accurately track the rapidly changing TPO when the timing frequency offset (TFO) and timing jitter offset (TJO) are present. In fact, the MM-TR is not suitable for processing TDMA burst signals.

Many efforts have been made to reduce TDMA overheads in 50G-PON. The DSP-assisted burst-mode receiver was proposed to achieve fast burst synchronization and equalization within ~200 ns for processing the 25Gb/s burst signal in 50G-PON, and was not involved in BM-TR \cite{zheng2019high}. Our previous work \cite{zhou2024burst} demonstrated BM-TR based on a frequency tone for the 50Gb/s PON based on discrete multi-tone, which has an excellent performance for estimating the rapidly changing TPO when the TFO and TJO are present. However, the frequency tone inserted in the spectrum obviously degrades the performance of the single-carrier OOK-based 50G-PON. The feed-forward squaring timing recovery algorithm was used to calculate the TPO fast for the BM-TR \cite{zhang2018real, zhang2021demonstration}. However, the feed-forward squaring timing recovery needs more than double-rate sampling, which does not meet the requirement for single-rate sampling in 50G-PON. A non-data-aided frequency-domain phase detector was proposed by Barton and Al-Jalili (BAJ) for implementing the feed-forward TR \cite{matalla2022real, matalla2021hardware}. However, the BAJ-PD requires more than single-rate sampling. Therefore, no TR exists that can fast estimate the varying TPO for the 1-SPS signal. 

\begin{figure*}[!tb]
\centering
\includegraphics[width=\linewidth]{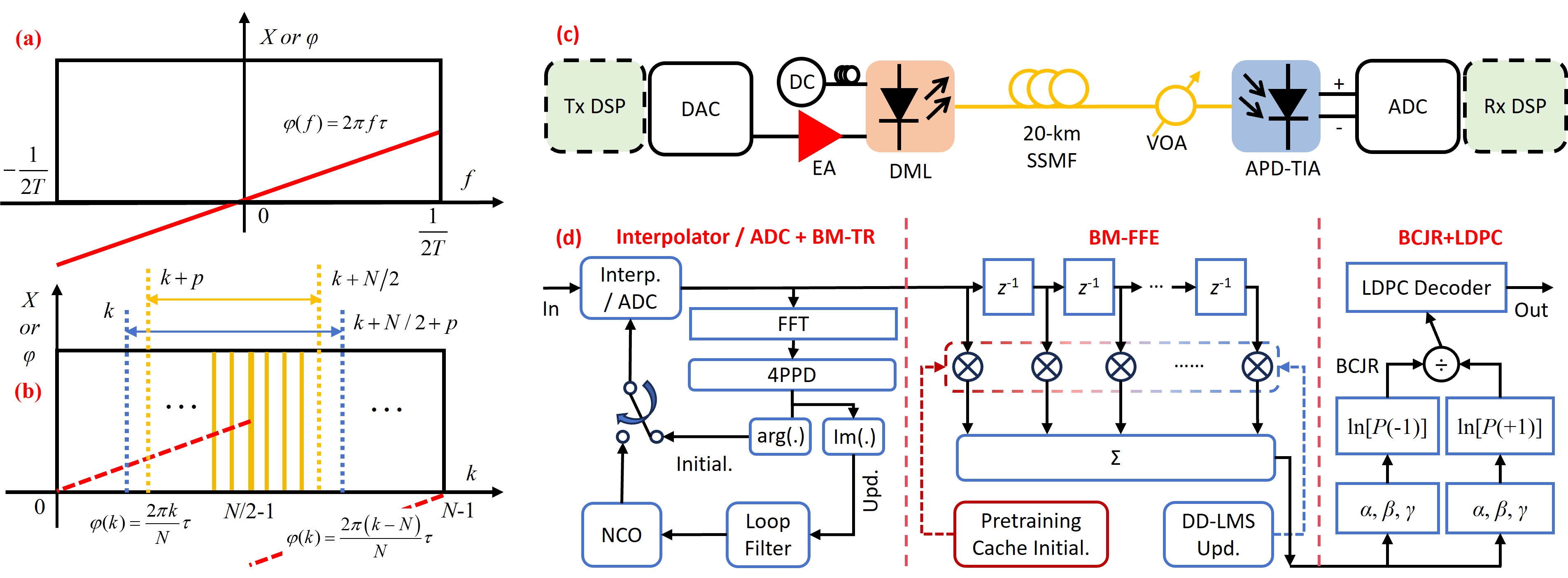}
\caption{(a) The signal spectrum (Black box) and phase-frequency response (Red solid line) where $T$ is the symbol duration. (b) The discrete signal spectrum (Yellow lines) and phase-frequency response (Red dashed line) for the 1-SPS discrete signal, where $N$ is the FFT size. (c) The experimental setups of the 50G-PON based on the DML and APD-TIA. (d) The BM-DSP architecture with the 4PPD-based BM-TR, BM-FFE, BCJR, and LDPC. Interp.: Interpolator. Initial.: Initialization. Upd.: Updating.}
\label{fig:2}
\end{figure*}

In this paper, we propose the first BM-TR based on a fourth-power phase detector (4PPD) without using training sequences for the 1-SPS burst signal in the 50G-PON. The proposed 4PPD can directly calculate the TPO, thereby initializing the BM-TR to avoid a long convergence time and reduce burst overhead. After the initialization, the feedback loop structure tracks the TPO changes based on the sign of 4PPD, loop filter, and NCO. The main contributions of this paper are as follows: 
\begin{itemize}
\item We propose the 4PPD to overcome the challenge of directly calculating the TPO for a 1-SPS OOK signal. In addition, the robustness and completeness of the 4PPD have been rigorously proven via mathematical derivation.
\item We establish an experiment of 50G-PON to verify the effectiveness of 4PPD-based BM-TR with timing offset initialization (TOI) under slowly-varying TPO as well as rapidly-changing TPO in the presence of TFO and TJO.
\end{itemize}

The remainder of this paper is organized as follows. The principle of 4PPD for the 1-SPS signal is shown in Section \ref{Section2}. In Section \ref{Section3}, the experimental setups and results for 50G-PON are demonstrated to verify the performance of 4PPD-based BM-TR. The paper is concluded in Section \ref{Section4}. 

\section{Principle of 4PPD for 1-SPS Signal} \label{Section2}
Here, we demonstrate the principle of 4PPD for the 1-SPS signal through mathematical derivation. When the TPO is applied to time-domain signal $a(t)$, its frequency-domain signal $X(f)$ can be expressed as
\begin{equation}
X(f) = \mathcal{F}\big\{h(t)\otimes a(t+\tau)\big\}=H(f)A(f)e^{j2\pi f\tau}
  \label{eq:1}
\end{equation}
where $A(f)$ is the frequency-domain signal of $a(t)$. $h(t)$ is the impulse response with $L$ taps and $H(f)$ is the frequency response of channel. $\tau$ is the true value of TPO. Fig. \ref{fig:2} (a) shows the signal spectrum and phase-frequency response, where $T$ is the sampling duration. The red solid line shows the phases caused by the TPO that vary linearly with frequency. The fast Fourier transform (FFT) transfers the time-domain digital received signal after the ADC to the discrete frequency-domain signal, which can be represented as
\begin{equation}
X(k) =
\begin{cases}
H(k)A(k)e^{j\frac{2\pi k}{N}\tau}, &\text{$0\leq k\leq N/2-1$},\\
H(k)A(k)e^{j\frac{2\pi (k-N)}{N}\tau}, &\text{$N/2\leq k\leq N-1$}
\end{cases}\label{eq:2}
\end{equation}
where $A(k)$ is the output of FFT for the transmitted signal $a(n)$. $H(k)$ is the FFT for the channel impulse response. $N$ is the size of the FFT. Fig. \ref{fig:2} (b) shows the discrete signal spectrum and phase-frequency response for the 1-SPS discrete signal, where $N$ is the FFT size. The red dashed line shows the discrete phases that vary linearly with discrete frequency points. The 4PPD aims to calculate the value of $\tau$ or estimate its trend. For the 4PPD, the fourth-power expression can be expressed as
\begin{equation}
\left[\sum^{N/2-1}_{k=0}X(k)X^{*}(N/2+k+p)\right]\left[\sum^{N/2-1}_{k=0}X^{*}(k+N/2)X(k+p)\right] \label{eq:3}
\end{equation}
where $X(N/2+k+p)$ and $X(k+p)$ are the cyclic extension of the frequency-domain signal. The range of the integer $P$ is from $-N/2$ to $N/2$. The first term of Eq. (\ref{eq:3}) can be extended as
\begin{equation}
\begin{aligned}
&\sum^{N/2-1}_{k=0} X(k)X^{*}(N/2+k+p)
\\= &\sum^{N/2-1}_{k=0} Y(k)e^{j2\pi\frac{k\tau}{N}}Y^{*}(N/2+k+p)e^{-j2\pi\frac{(-N/2+k+p)\tau}{N}}
\\= &\sum^{N/2-1}_{k=0} Y(k)Y^{*}(N/2+k+p)e^{j2\pi\frac{(N/2-p)\tau}{N}}
    \label{eq:4}
\end{aligned}
\end{equation}
where $Y(k) = H(k)A(k)$. The second term of Eq. (\ref{eq:3}) can be extended as
\begin{equation}
\begin{aligned}
&\sum^{N/2-1}_{k=0} X^{*}(k+N/2)X(k+p)
\\= &\sum^{N/2-1}_{k=0} Y^*(k+N/2)e^{-j2\pi\frac{(k-N/2)\tau}{N}}Y(k+p)e^{j2\pi\frac{(k+p)\tau}{N}}
\\= &\sum^{N/2-1}_{k=0} Y^{*}(k+N/2)Y(k+p)e^{j2\pi\frac{(N/2+p)\tau}{N}}.
    \label{eq:5}
\end{aligned}
\end{equation}

\begin{figure*}[!tb]
\centering
\includegraphics[width=\linewidth]{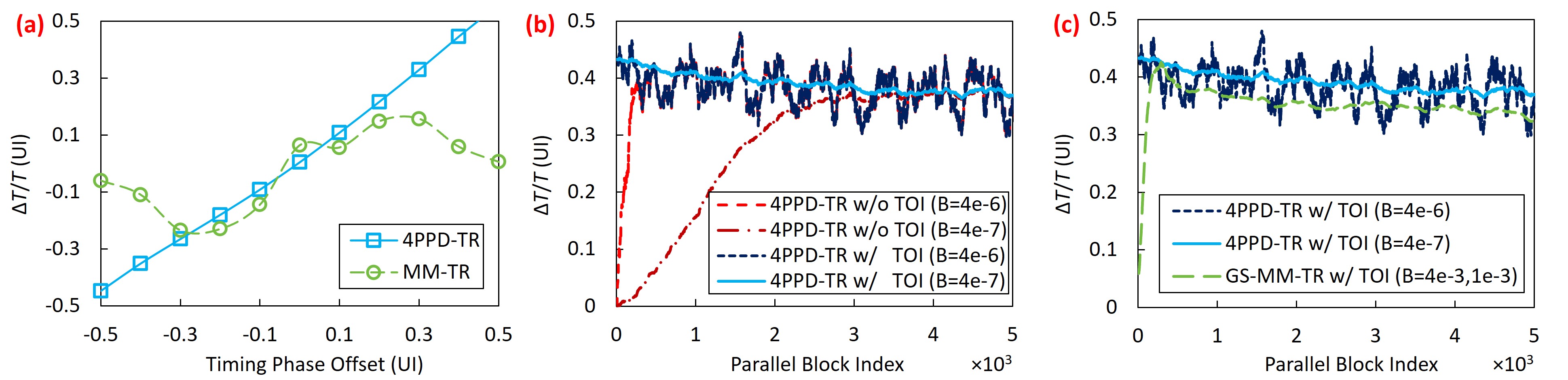}
\caption{(a) The timing offset initialization (TOI) using 4PPD and MM-PD. (b) The slowly-changing TPO estimated by the 4PPD-based BM-TR with TOI and 4PPD-based TR without TOI when the noise bandwidth $B$ is set to $4\times 10^{-6}$ or $4\times 10^{-7}$, respectively. (c) TPO estimated by the 4PPD-based BM-TR with TOI and the GS-MM-TR using a noise bandwidth $B$ of $5\times 10^{-3}$ initially for faster convergence, then switching to $1\times 10^{-3}$ for steady-state tracking.}
\label{fig:3}
\end{figure*}

Next, we will calculate Eq. (\ref{eq:3}) in detail. When the transmitted signal $a(n) \in \{\pm 1\}$ (i.e., PAM2), its output after the FFT satisfies the Hermitian conjugate property, $Y(k) = Y^{*}(N-k)$, where $k$ is from 1 to $N/2-1$. Meanwhile, $Y(0)$ and $Y(N/2)$ are real values. We define $b(n) = (-1)^{n}y(n)e^{-j\frac{2\pi}{N}np}$ where $y(n)$ is the corresponding time-domain signal of $Y(k)$. After FFT, the $B(k)$ can be defined as $Y(N/2 + k + p)$ where $k$ and $n$ are from $0$ to $N-1$, respectively. Based on Parseval's theorem, the frequency-domain correlations can be calculated by
\begin{equation}
\begin{aligned}
    S_H =\sum^{N-1}_{k=0} Y(k)B^{*}(k)&= N \sum^{N-1}_{n=0} y(n)\left[(-1)^{n}y(n)e^{j\frac{2\pi}{N}np}\right]\\
    &= N \sum_{n=0}^{N-1}  (-1)^n e^{j\frac{2\pi p n}{N}}|y(n)|^2.
    \label{eq:8}
\end{aligned}
\end{equation}
The signal $y(n)$ is decomposed into an ideal signal $a(n)$ with the inter-symbol interference (ISI) $\varepsilon(n)$ as
\begin{equation}
y(n) = a(n) + \varepsilon(n)
\label{pn}
\end{equation}
where $\varepsilon(n)= \sum_{l=1}^{L-1} h(l)a(n-l)$. Hence, by substituting Eq. (\ref{pn}) into Eq. (\ref{eq:8}), we can obtain that
\begin{equation}
\begin{aligned}
S_H &= N \sum_{n=0}^{N-1} (-1)^n e^{j\frac{2\pi p n}{N}} \left[|a(n)|^2+ 2a(n)\varepsilon(n)+ |\varepsilon(n)|^2 \right]\\&=N \sum_{n=0}^{N-1} (-1)^n e^{j\frac{2\pi p n}{N}} |a(n)|^2 \left[1+2\frac{\varepsilon(n)}{a(n)}+ \frac{|\varepsilon(n)|^2}{|a(n)|^2}\right].
    \label{S_H}
\end{aligned}
\end{equation}
Moreover, the first term in Eq. (\ref{S_H}) can be derived that
\begin{equation}
N\sum^{N-1}_{n=0} (-1)^{n}e^{j\frac{2\pi}{N}np}|a(n)|^2 = N \frac{1-e^{j\pi(N+2p)}}{1-e^{j\pi(1+\frac{2p}{N})}}=0.
\end{equation}
For the TR, ISI affects the performance of any phase detection. Typically, the limited bandwidth can be addressed by using an analog continuous-time linear equalizer or a short-tap FFE in the TR loop, thereby reducing ISI. The $\varepsilon(n)$ is sufficiently small relative to the signal (i.e., $|\varepsilon(n)| \ll |a(n)|$), which results in $S_H \approx 0$. Thus, we can mathematically verify the 4PPD in the ideal case. Moreover, $S_H$ in Eq. (\ref{eq:8}) can be decomposed into two summation terms as
\begin{equation}
\begin{aligned}
    &\sum_{k=0}^{N-1} Y(k)Y^*(N/2+k+p)\\ = &\left\{ \sum_{k=0}^{N/2-1} + \sum_{k=N/2}^{N-1} \right\} Y(k)Y^*(N/2+k+p) \\
=& \underbrace{\sum_{k=0}^{N/2-1} Y(k)Y^*(N/2+k+p)}_{L(p)} + \underbrace{\sum_{k=0}^{N/2-1} Y(k+N/2)Y^*(k+p)}_{R(p)}.
\label{eq:9}
\end{aligned}
\end{equation}
Then, it can be derived that
\begin{equation}
L(p) \cdot R^*(p) = -|L(p)|^2 + S_H^* \times L(p) \approx -|L(p)|^2.
\label{LR*}
\end{equation}
Based on Eqs. (\ref{eq:4}) and (\ref{eq:5}), the fourth-power expression in Eq. (\ref{eq:3}) can be rewritten as
\begin{equation}
\begin{aligned}
&\left[\sum^{N/2-1}_{k=0}X(k)X^{*}(N/2+k+p)\right]\left[\sum^{N/2-1}_{k=0}X^{*}(k+N/2)X(k+p)\right]\\
&=L(p)\times R^*(p) \times e^{j2\pi \tau}.
    \label{LR*exp}
\end{aligned}
\end{equation}
By substituting Eq. (\ref{LR*}) into Eq. (\ref{LR*exp}), the 4PPD can directly calculate the $\tilde{\tau}(p)$ as
\begin{equation}
\tilde{\tau}(p) = \frac{\text{arg}\{L(p)\times R^*(p)\times e^{j2\pi \tau}\}}{2\pi}-0.5.
\label{tau0}
\end{equation}
When $|\varepsilon(n)| \ll |a(n)|$, the estimated TPO $\tilde{\tau}(p)$ is approximately equal to the true value of TPO $\tau$.To improve the accuracy, the average operation can be implemented by
\begin{equation}
\overline{\tau} = \frac{\text{arg}\left\{\sum_{p=-N/M}^{N/M}L(p)\times R^*(p) \times e^{j2\pi \tau}\right\}}{2\pi}-0.5,
\label{eq:13}
\end{equation}
which can calculate a more accurate value of TPO for initializing the BM-TR or feed-forward TR. After the initialization, the sign of Eq. (\ref{tau0}) with $p=0$ can be used to estimate the trend of $\tau$ for the timing recovery with a feedback loop structure, which can be expressed as
\begin{equation}
\begin{aligned}
OUT_{\text{4PPD}} &= \text{Im}\left[L(0)\times R^*(0) \times e^{j2\pi \tau}\right]\\
& \approx -|L(0)|^2 \sin(2\pi\tau)
    \label{OUT4}
\end{aligned}
\end{equation}
where $\text{Im}(\cdot)$ denotes the imaginary part of a complex number. 

In conclusion, Eq. (\ref{S_H}) confirms that the 4PPD is robust to mild ISI in 50G PON. The noise does not introduce additional bias but increases the estimation variance. For implementing the BM-TR, the TPO can be directly calculated by Eq. (\ref{eq:13}) using the payload, which does not increase the overhead. Then, the estimated TPO initializes the BM-TR to achieve fast convergence. After the initialization, the feedback loop structure based on the trend of TPO from Eq. (\ref{OUT4}), loop filter, and NCO tracks the TPO. In fact, the loop filter in the feedback loop structure can average the trend of TPO to improve the convergence accuracy. 

\section{Experimental Setups and Results} \label{Section3}
An experiment of 50G-PON was set up to verify the feasibility of the 4PPD-based BM-TR. Fig. \ref{fig:2} (c) shows the experimental setups of the uplink for the 50G-PON based on the directly modulated laser (DML) and the avalanche photo-diode with a trans-impedance amplifier (APD-TIA). Firstly, the DSP at the transmitter (Tx DSP) generated the two-level pulse-amplitude modulation (PAM2) signal. The digital PAM2 signal was converted to an analog signal by a digital-to-analog converter (DAC). An electrical amplifier amplified the analog signal. Then, the amplified analog signal was modulated on an O-band optical carrier by a 25G-class transmitter optical subassembly (TOSA) with a DML. The optical OOK signal was then launched into a 20km standard single-mode fiber (SSMF). A variable optical attenuator (VOA) was used to adjust the received optical power (ROP). A 25G-class receiver optical subassembly (ROSA) using the APD-TIA was used to convert the optical signal to an electrical signal. The attenuation of the signal at the Nyquist frequency is less than 20dB. The electrical signal was then sent to an analog-to-digital converter (ADC) to convert the analog signal to a digital signal. In the offline experiment, an interpolator was used to simulate the physical ADC. Finally, the DSP at the receiver (Rx DSP) was used to recover the digital signal.

\begin{figure}[!tb]
\centering
\includegraphics[width=0.9\linewidth]{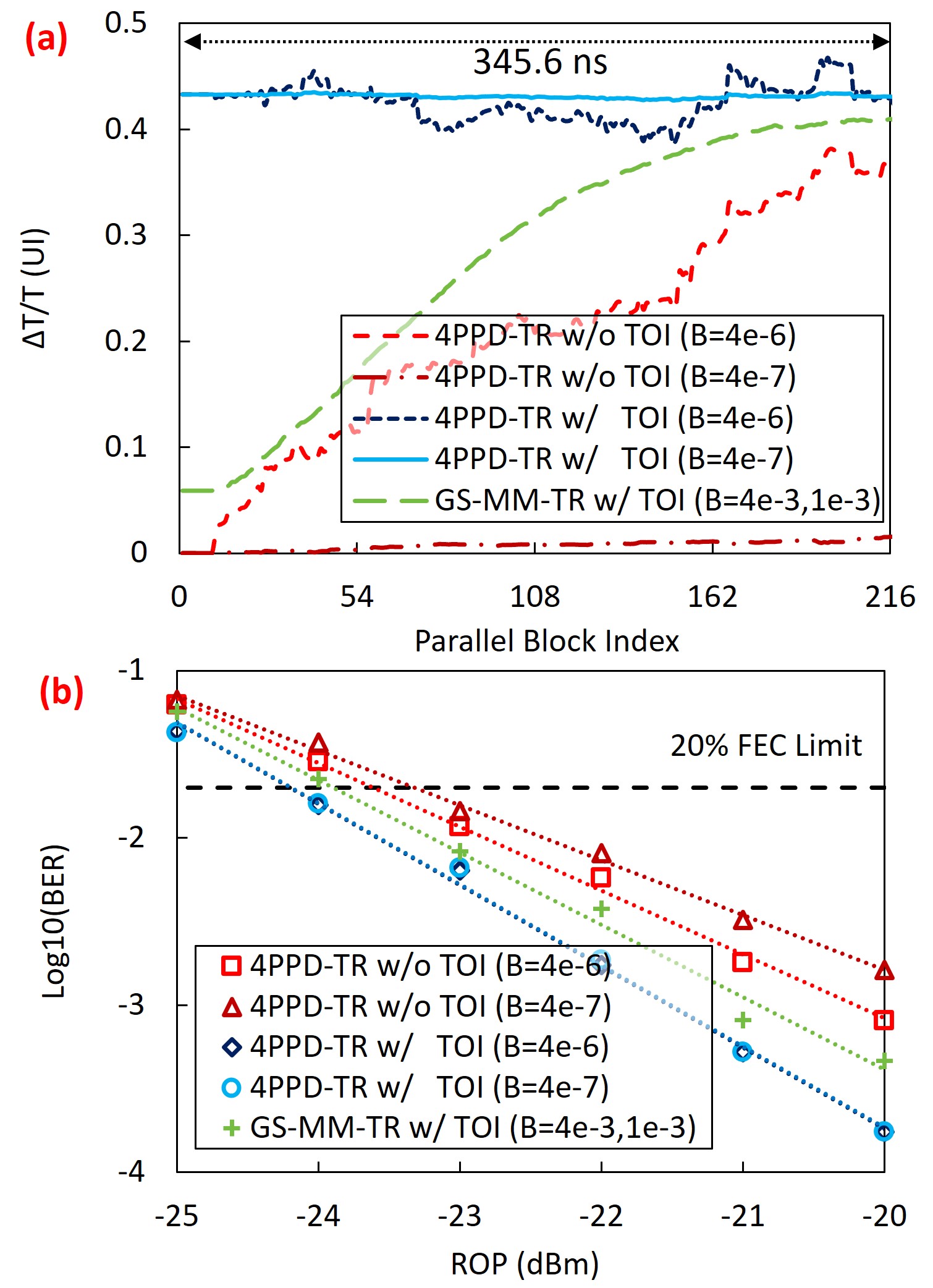}
\caption{(a) The slowly-changing TPO of the first LDPC block. (b) BER versus ROP of the first LDPC block by the 4PPD-based BM-TR with TOI or without TOI, and GS-MM-TR with TOI.}
\label{fig:4}
\end{figure}

\begin{figure*}[!tb]
\centering
\includegraphics[width=\linewidth]{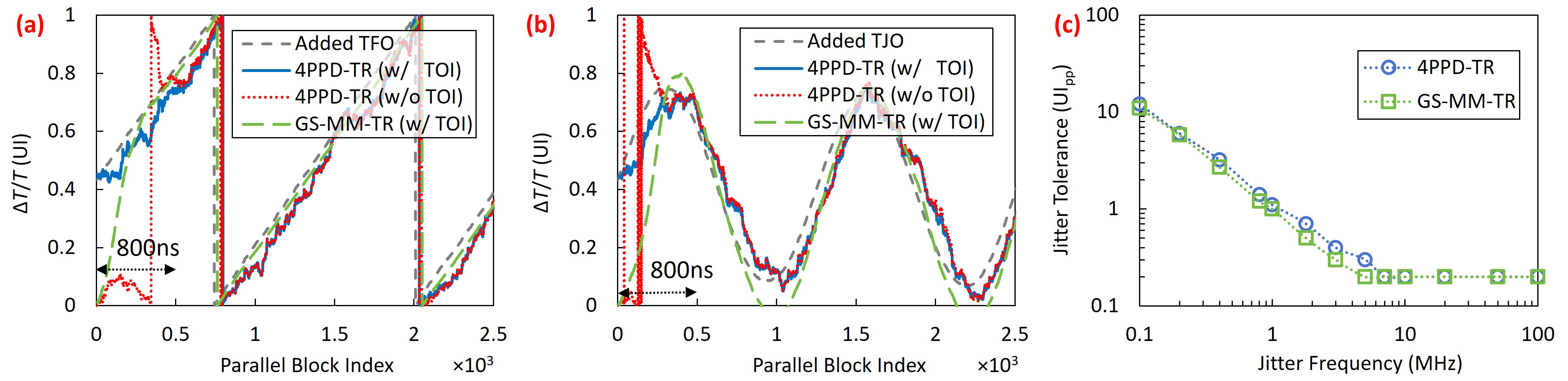}
\caption{The rapidly-changing TPO estimated by the 4PPD-based BM-TR with TOI or without TOI, and the GS-MM-TR with TOI under (a) TFO and (b) TJO, respectively. (c) Jitter tolerance of the 4PPD-based BM-TR and GS-MM-TR.}
\label{fig:5}
\end{figure*}

Fig. \ref{fig:2} (d) shows the BM-DSP architecture with the 4PPD-based BM-TR, burst-mode feedforward equalizer (BM-FFE) with 21 taps, Bahl-Cocke-Jelinek-Raviv (BCJR) for soft-output detection, and low-density parity-check (LDPC) with a code rate of about 5/6 and a code word of 17280 bits. The BM-TR is positioned in front of the FFE, as it allows the timing recovery to operate on the raw signal. This is necessary because the FFE requires a properly sampled signal to function effectively. Despite the presence of ISI, the 4PPD can still extract a sufficiently accurate TPO for initialization. In the 4PPD-based BM-TR, the digital signal is fed into a 112-point FFT, containing 80 payload symbols and 32 overlapping symbols. It is determined by the parallelism degree of the data path (i.e., the clock is assumed to be 625 MHz). The 112-point FFT can be efficiently implemented using a mixed-radix FFT (i.e., 4 radix-2 layers and one radix-7 layer) \cite{zhou2025real}. After the FFT, the TPO calculated by the 4PPD first initializes the BM-TR, enabling fast convergence. After the initialization, the BM-TR switches to the feedback loop structure based on the trend of the 4PPD, loop filter, and NCO. The loop delay of $\sim$20 cycles has been accounted for in the DSP model.

Figure \ref{fig:3} (a) shows the estimated timing offset initialization (TOI) using 4PPD and MM-PD. The timing offset estimated by 4PPD is linearly related to the pre-set value, while MM-PD fails to accurately estimate it without the feedback loop. Fig. \ref{fig:3} (b) shows the estimated TPO by the 4PPD-based BM-TR with TOI and 4PPD-based TR without TOI under the slowly-changing TPO when the noise bandwidth $B$ is set to $4\times 10^{-6}$ or $4\times 10^{-7}$, respectively. A larger $B$ leads to faster convergence but larger fluctuation in the estimated TPO; a smaller $B$ results in slower convergence but reduced fluctuation. When the $B$ is set to $4\times 10^{-7}$, the 4PPD-based TR without TOI requires 3000 parallel blocks (i.e., $3000\times80 = 240000$ symbols) to converge to the right TPO. When the $B$ is set to $4\times 10^{-6}$, the 4PPD-based TR without TOI requires more than 300 parallel blocks (i.e., 24000 symbols) to converge, which is still overlong. Meanwhile, it also leads to a 0.1 UI fluctuation. Fig. \ref{fig:3} (c) shows the TPO estimated by the 4PPD-based BM-TR with TOI and the gear-shifting (GS) MM-TR using a noise bandwidth $B$ of $5\times 10^{-3}$ initially for faster convergence, then switching to $1\times 10^{-3}$ for steady-state tracking. GS-MM-TR requires about 300 blocks to converge. Due to the influence of the channel, the difference in the timing offset of their converging tracking is $\sim$0.05 UI. Therefore, only 4PPD-based BM-TR with TOI can track the slowly-changing TPO rapidly, almost without convergence time. Meanwhile, we can use the $B$ of $4\times 10^{-7}$ for the 4PPD-based BM-TR with TOI to achieve a stable estimated TPO. 

Figure \ref{fig:4} (a) depicts the estimated TPO of the first LDPC block by the 4PPD-based BM-TR with TOI and 4PPD-based TR without TOI under the slowly-changing TPO. The first 216 parallel blocks correspond to the first LDPC block with 17280 bits. The 4PPD-based BM-TR with TOI can fast converge to the right TPO within the first LDPC block. However, both the 4PPD-based TR without TOI and the GS-MM-TR with an inaccurate TOI fail to converge to the correct TPO within the first LDPC block. The signal within the tracking time has worse performance than that within the converged time. Therefore, the required ROP at the 20\% FEC limit of the first LDPC block usually determines the receiving sensitivity. Fig. \ref{fig:4} (b) shows the BER versus ROP of the first LDPC block by the 4PPD-based BM-TR with TOI or without TOI, and GS-MM-TR with TOI under the slowly-changing TPO. The 4PPD-based TR without TOI under the $B$ of $4\times 10^{-6}$ has better BER performance than that under the $B$ of $4\times 10^{-7}$. GS-MM-TR with TOI has better BER performance than the 4PPD-based TR without TOI, while the 4PPD-based BM-TR with TOI under the $B$ of $4\times 10^{-6}$ or $4\times 10^{-7}$ has the best BER performance. Due to the almost zero convergence time, the receiving sensitivity using 4PPD-based BM-TR with TOI is $0.5$dB higher than that using 4PPD-based BM-TR without TOI.

Figure \ref{fig:5} depicts the rapidly-changing TPO estimated by the e 4PPD-based BM-TR with TOI or without TOI, and the GS-MM-TR with TOI under (a) TFO and (b) TJO, respectively. To track the rapidly-changing TPO, $B$ is set to $4\times 10^{-6}$ for both the 4PPD-based BM-TR with TOI or without TOI. The 4PPD-based BM-TR with TOI can track the fast-changing TPO almost without convergence time under TFO and TJO, respectively. Due to the loop delay, there is a fixed delay on the estimated TPO compared to the added TPO. The 4PPD-based TR without TOI requires approximately 500 parallel blocks (i.e., 40000 symbols, 800ns) to track the right TPO under (a) TFO and 400 parallel blocks (i.e., 32000 symbols, 640ns) for (b) TJO, respectively. GS-MM-TR requires about 300 parallel blocks (i.e., 24000 symbols, 480ns) to converge. Obviously, the 4PPD-based TR without TOI or GS-MM-TR with TOI consumes more overhead than the 4PPD-based BM-TR. Therefore, 4PPD-based BM-TR has better performance for processing the burst signal with TFO than 4PPD-based TR without TOI and GS-MM-TR with TOI. Fig. \ref{fig:5} (c) shows the jitter tolerance of the 4PPD-based BM-TR and GS-MM-TR. Similar to MM-TR, the 4PPD-based TR tolerates higher than about 1 $\mathrm{UI_{pp}}$ of sinusoidal jitter at frequencies below 1 MHz.

\begin{figure}[!tb]
\centering
\includegraphics[width=0.9\linewidth]{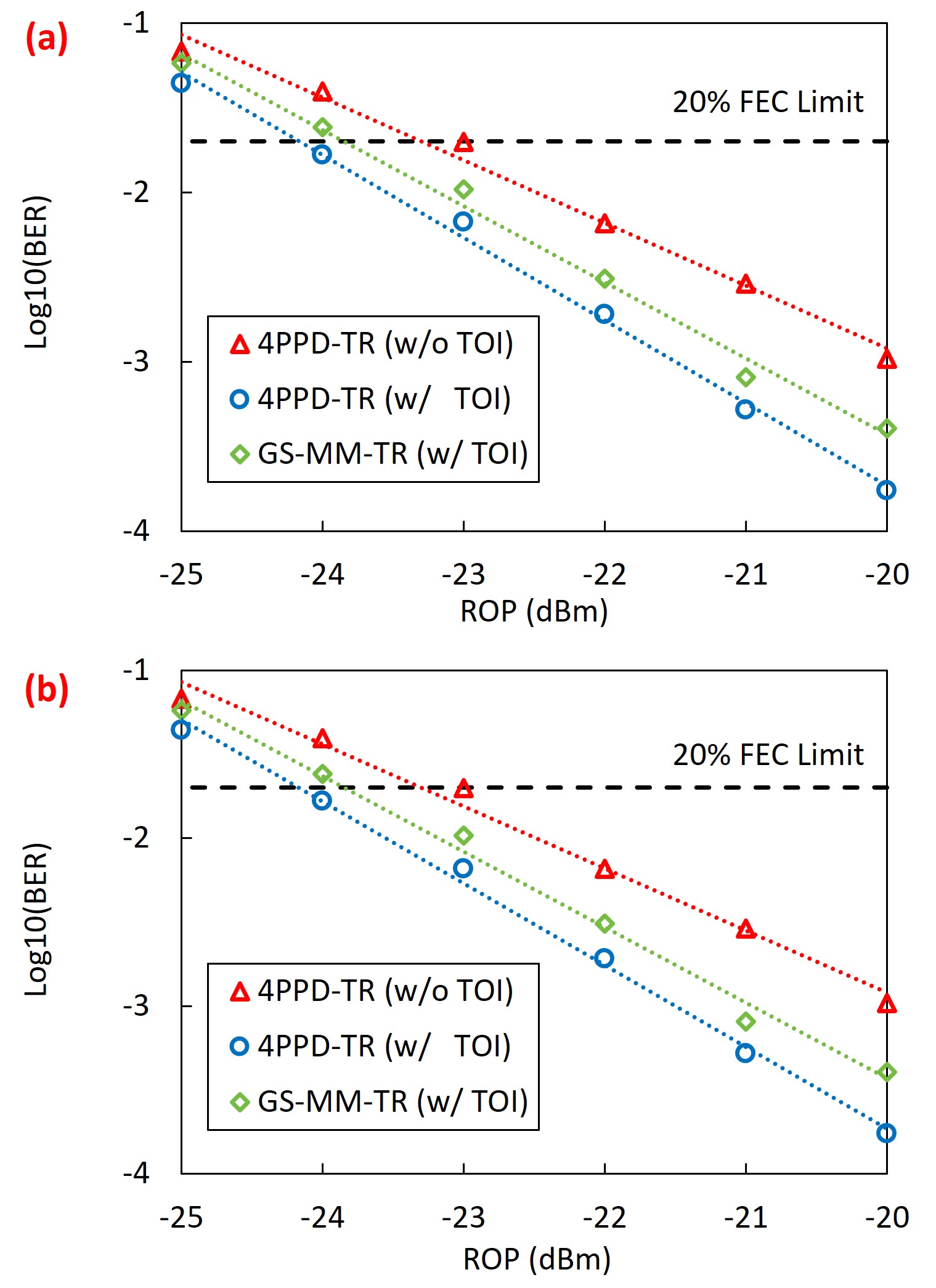}
\caption{BER versus ROP of the first LDPC block by the 4PPD-based BM-TR with TOI or without TOI, and GS-MM-TR under (a) TFO and (b) TJO, respectively.}
\label{fig:6}
\end{figure}

Figure \ref{fig:6} shows the BER versus ROP of the first LDPC block by the 4PPD-based BM-TR with TOI or without TOI, and GS-MM-TR under (a) TFO and (b) TJO, respectively. The $B$ is set to $4\times 10^{-6}$ for the 4PPD-based BM-TR. The 4PPD-based TR without TOI cannot converge to the right TPO within the first LDPC block, which seriously degrades the BER performance of the first LDPC block. The BER performance of GS-MM-TR with TOI is better than that of the 4PPD-based BM-TR without TOI. Compared to the 4PPD-based TR without TOI, 4PPD-based BM-TR with TOI leads to an approximately $1$dB enhancement in receiver sensitivity under both the TFO and TJO. The receiver sensitivity using 4PPD-based BM-TR can achieve $-24$dBm under rapidly-changing TPO in the presence of TFO and TJO, which is almost similar to that under slowly-varying TPO. This result indicates that 4PPD-based BM-TR with TOI has a good performance in compensating for rapidly-changing TPO.

Although in terms of convergence speed, 4PPD-based BM-TR is superior to MM-TR, the computational complexity of 4PPD is higher than that of MMPD. At the initialization stage, there are 8768 real-value multiplications and 9663 additions for 4PPD, including the implementation of a 112-point FFT. At the tracking stage, there are 1505 real-value multiplications and 2526 additions for 4PPD, including the implementation of a 112-point FFT. Since the overlapping is not required, there are 160 real-value multiplications and 159 additions for MM-PD.

\section{Conclusions} \label{Section4}
In this paper, we propose the 4PPD to overcome the challenge of directly calculating the TPO from a 1-SPS OOK signal. The robustness and completeness of the 4PPD have been rigorously proven via mathematical derivation. The calculated TPO initializes the 4PPD-based BM-TR. Owing to TPO initialization, the 4PPD-based BM-TR avoids the long convergence times and eliminates the burst overhead within 20 cycles, while MM-PD fails to accurately estimate it without the feedback loop. We establish an experiment of 50G-PON to verify the effectiveness of 4PPD-based BM-TR under slowly-varying TPO as well as rapidly-changing TPO in the presence of TFO and TJO. Compared to 4PPD-based TR without TOI, 4PPD-based BM-TR with TOI leads to approximately $0.5$ dB and $1$dB improvement in receiver sensitivity under slowly-varying and rapidly-changing TPO, respectively. Although 4PPD-based BM-TR converges faster than MM-TR, 4PPD has higher computational complexity than MMPD. Furthermore, the feedforward TR based on the calculated TPO can avoid the 20 cycles burst overhead at the cost of performance. In conclusion, 4PPD-based BM-TR proves to be particularly effective in handling 1-SPS burst signal for 50G-PON.

\section{Acknowledgements}
This work was supported in part by the National Key R\&D Program of China under Grant 2023YFB2905700, in part by the National Natural Science Foundation of China under Grant 62371207 and Grant 62005102, in part by the Young Elite Scientists Sponsorship Program by CAST under Grant 2023QNRC001, and in part by the Hong Kong Research Grants Council GRF under Grant 15231923, in part by the Zhuhai Industrial Core and Key Technology Research and Development Project under Grant 2320004002531, in part by the Guangdong Province Cross domain intelligent detection and information processing innovation team under Grant 2023KCXTD044, in part by the Guangdong Province Key Laboratory of Intelligent Detection in Complex Environment of Aerospace, Land and Sea under Grant 2022KSYS016.

\bibliography{sample}
\bibliographyfullrefs{sample}

\end{document}